\documentclass[preprint,12pt, nopreprintline]{elsarticle}

\usepackage{amssymb}
\usepackage{amsmath}
\usepackage{xcolor}
\usepackage{multirow}
\usepackage{graphicx}
\usepackage{float}
\usepackage{pdfpages}
\usepackage{pifont}
\usepackage{hyperref}
\usepackage{tabularx}
\usepackage{longtable}
\usepackage{xltabular}

\begin{document}

\begin{frontmatter}



\title{Dual Attention Model with Reinforcement Learning for Classification of Histology Whole-Slide Images}


\author[TIA]{Manahil Raza} 
\ead{manahil.raza@warwick.ac.uk}
\author[TIA]{Ruqayya Awan}
\ead{ruqayya.awan@warwick.ac.uk}
\author[TIA]{Raja Muhammad Saad Bashir}
\ead{saad.bashir@warwick.ac.uk}
\author[TIA]{Talha Qaiser}
\ead{talha.qaiser@warwick.ac.uk}
\author[TIA,ALAN]{Nasir M. Rajpoot}
\ead{n.m.rajpoot@warwick.ac.uk}

\affiliation[TIA]{organization={Tissue Image Analytics Centre, University of Warwick},
            city={Coventry},
            country={United Kingdom}}

\affiliation[ALAN]{organization={The Alan Turing Institute},
            city={London},
            country={United Kingdom}}

\begin{abstract}
Digital whole slide images (WSIs) are generally captured at microscopic resolution and encompass extensive spatial data (several billions of pixels per image). Directly feeding these images to deep learning models is computationally intractable due to memory constraints, while downsampling the WSIs risks incurring information loss. Alternatively, splitting the WSIs into smaller patches (or tiles) may result in a loss of important contextual information. In this paper, we propose a novel dual attention approach, consisting of two main components, both inspired by the visual examination process of a pathologist: The first {\em soft} attention model processes a low magnification view of the WSI to identify relevant regions of interest (ROIs), followed by a custom sampling method to extract diverse and spatially distinct image tiles from the selected ROIs. The second component, the {\em hard} attention classification model further extracts a sequence of multi-resolution glimpses from each tile for classification. Since hard attention is non-differentiable, we train this component using reinforcement learning to predict the location of the glimpses.  This approach allows the model to focus on essential regions instead of processing the entire tile, thereby aligning with a pathologist's way of diagnosis. The two components are trained in an end-to-end fashion using a joint loss function to demonstrate the efficacy of the model. The proposed model was evaluated on two WSI-level classification problems: Human epidermal growth factor receptor 2 (HER2) scoring on breast cancer histology images and prediction of Intact/Loss status of two Mismatch Repair (MMR) biomarkers from colorectal cancer histology images. We show that the proposed model achieves performance better than or comparable to the state-of-the-art methods while processing less than 10\% of the WSI at the highest magnification and reducing the time required to infer the WSI-level label by more than 75\%. The code is available at \href{https://github.com/manahilr/Dual-Attention}{github}.

\end{abstract}





\end{frontmatter}



\section{Introduction}

The current \emph{gold standard} in clinical practice for cancer diagnostics is the visual assessment of glass tissue slides by a pathologist \cite{de2010immunohistochemistry}. However, this process is tedious, time-consuming and subject to inter- and intra- observer variability \cite{codipilly2023evolving}. Digital slide scanners have enabled glass tissue slides to be scanned into digitized multi-gigapixel whole slide images (WSIs). Computational Pathology (CPath) models, particularly deep learning (DL) based models, can enhance precision and objectivity by quantitatively analyzing WSIs and avoiding human bias \cite{niazi2019digital,hosseini2023computational, zhou2020comprehensive,jahanifar2023domain}. In recent years, there has been growing interest in applying deep learning techniques to CPath problems. Algorithms solving tasks such as cancer grading \cite{qaiser2019learning}, tissue type classification \cite{91_abbet2022cpath,raza2023stain} and tumor segmentation \cite{dawood2022cellular,graham2019hover} have received considerable attention within the CPath community. However, despite these notable advancements, there remain several challenges associated with the automated processing of WSIs. 

Whole slide images are labeled at the slide-level and each slide can potentially contain hundreds or even thousands of image tiles or patches. Slide-level labels often correspond to a small region within the large gigapixel image, making it difficult to determine which specific region in the WSI contributed to the overall label \cite{gadermayr2024multiple}. Moreover, DL-based CPath methods require comprehensive data to be trained effectively and thus rely on either large datasets with slide-level labels in weakly-supervised environments or manually annotated WSIs in fully-supervised environments \cite{lu2021data}. However, obtaining manual patch or region of interest (ROI) level annotations is costly, time-consuming and labor intensive \cite{tang2023multiple}. Weakly-supervised approaches attempt to mitigate the annotation burden by assigning the slide-level labels to every patch. While this approach simplifies the labeling process, it can introduce significant noise into the training data, especially in cases where the tumor region may be localized to a small part of the whole slide image.  

Moreover, the inherent characteristics of WSIs, including their gigapixel size, varying magnification levels, and high-resolution, present additional challenges for automated processing in CPath workflows \cite{hosseini2023computational}. An uncompressed WSI may consume up to 46.9 gigabyte (GB) of memory \cite{holzinger2017towards}. Consequently, the computational and memory demands needed for the direct processing of multi-gigapixel WSIs exceed the capabilities of current hardware devices \cite{dimitriou2019deep}. A straightforward approach is to downsample the WSI, but this may result in the loss of crucial information required for various downstream tasks \cite{tripathi2021end}. An approach frequently employed in CPath literature involves the extraction of image tiles or patches from the tissue or tumor regions of the WSI using the ``sliding window'' paradigm, where each image patch is subsequently processed by the deep learning model \cite{qaiser2019learning,sanyal2021dan}. This method has two main disadvantages: first, not all patches are diagnostically relevant e.g., patches extracted from the background region or adipose tissue; second, patches extracted from the same spatial region can potentially provide redundant information to the model \cite{zhang2021joint}. Consequently, various patch-selection strategies and attention-based methods have been proposed to address these issues, by extracting patches based on specific selection criterion or mechanisms \cite{rodner2017deep, yang2019guided, xu2019attention, kong2021efficient}. 

Immunohistochemistry (IHC) is a variant of immunostaining which is widely used to detect and quantify the presence of specific protein markers in histopathology images \cite{taylor2006quantification}. IHC introduces an additional layer of complexity for automated WSI processing methods due to staining inconsistencies and variability  \cite{otali2016standard,nguyen2022immunohistochemistry}. Therefore, there is a need for a more robust and generalizable method. 

To determine the best strategy for quantitatively assessing WSIs, we first need to understand a pathologist's diagnostic process \cite{chakraborty2024decoding}. In 1967, King \emph{et al.} \cite{king1967does} raised the question, ``How does a pathologist make a diagnosis?''. This was first addressed by Pena \emph{et al.} \cite{pena2009does} who suggested that rather than using an exhaustive strategy wherein all available data is examined and evaluated - analogous to the sliding window paradigm in computational pathology - an experienced pathologist may use certain ``shortcuts'' by employing cognitive processes (such as saccades) to gather information from the given histopathology slide. Moreover, pathologists often examine the entire slide for a preliminary diagnosis, unlike traditional DL models trained on patches that primarily focus on local features rather than global features and risk incurring information loss \cite{zhou2021histopathology}.
By incorporating the shortcut approach used by a pathologist in IHC scoring or classification algorithms, we mimick the way a pathologist operates and consequently make the diagnostic process more understandable to the pathologists.

In this paper, we introduce a novel dual attention model that identifies, sequentially predicts, and analyzes diagnostically relevant regions in whole slide images (WSIs) using hierarchical attention and reinforcement learning. Inspired by the diagnostic process of a pathologist, this approach addresses the disjoint patching dilemma, which often leads to degraded visual context and loss of crucial information necessary for understanding the spatial organization of tissue components within the tumor microenvironment (TME). The model employs a soft attention module at low magnification to identify regions of interest (ROIs), analogous to a pathologist's initial scan. An attention sampling process is used to extract informative and spatially distinct high-power fields (HPFs) or image tiles from these ROIs. Subsequently, a hard attention module sequentially extracts and analyzes patches, ``glimpses" from these image tiles, similar to a pathologist's diagnostic saccades. Our end-to-end workflow includes a joint loss function to train both attention components simultaneously. We validate our method on two WSI slide-level classification tasks: predicting the Intact/Loss status of two Mismatch Repair (MMR) biomarkers and predicting human epidermal growth factor receptor 2 (HER2) scores [0-3+] for IHC-stained colorectal and breast cancer WSIs.

\begin{figure}[ht]
\centering
\includegraphics[width=\linewidth]{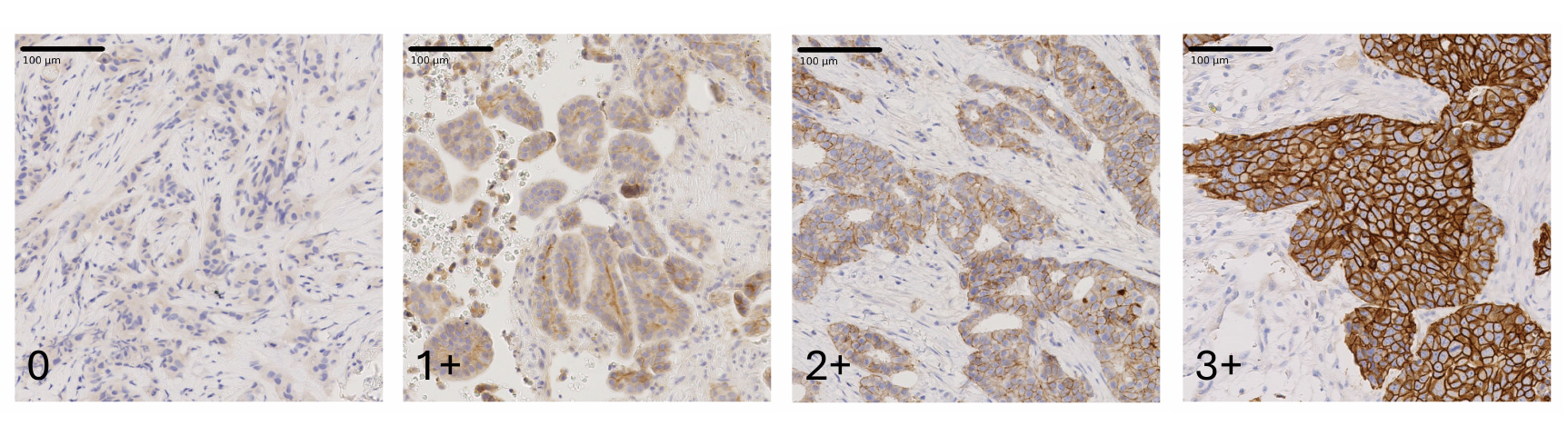}
\caption{Examples of regions of interest from breast cancer whole slide images with their respective HER2 Scores from the training dataset.}
\label{her2}
\end{figure}

The HER2 gene is a key cell membrane biomarker, over-expressed in approximately 25\% of invasive breast carcinomas and linked to high recurrence and poor survival rates  \cite{dean2008her2, hudis2007trastuzumab}. Anti-HER2 treatments have improved prognosis, and early diagnosis can reduce mortality by 40\% \cite{murthy2020tucatinib, wang2023weakly}. Current clinical guidelines recommend that all patients diagnosed with breast cancer undergo HER2 testing, using IHC staining, scored by pathologists from 0-3+  \cite{rakha2023uk, qaiser2018her}. However, approximately 20\% of HER2 tests are inaccurate, highlighting a need for improved or automated HER2 scoring methods \cite{bernasconi2012genetic}. Fig. \ref{her2} contains examples of ROIs from slides with different HER2 scores.

\begin{figure}[ht]
\centering
\includegraphics[width=\linewidth]{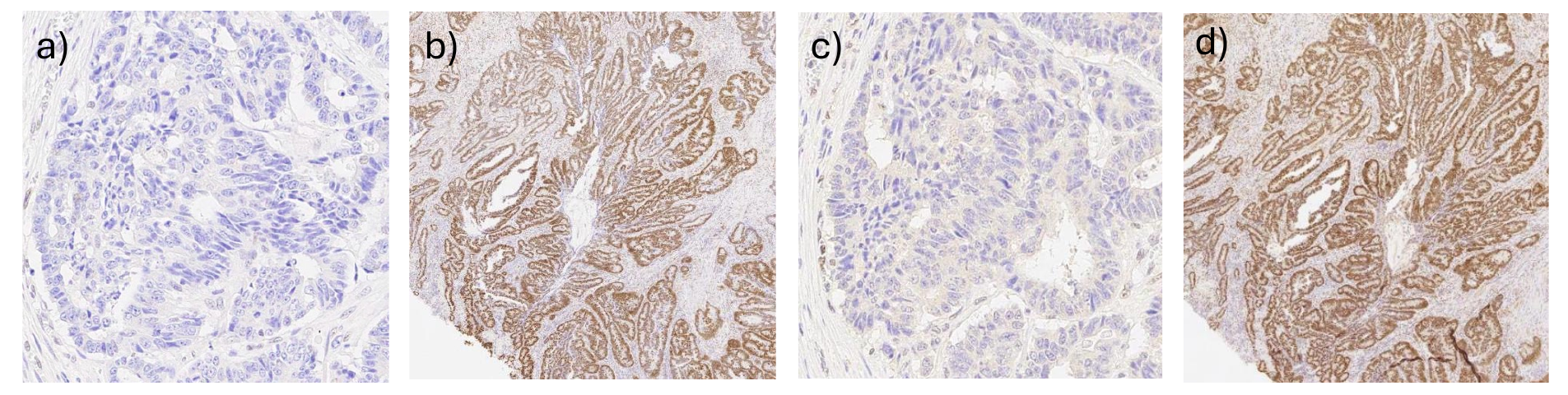}
\caption{Examples of regions of interest from colorectal cancer whole slide images: a) Loss status of MLH1, b) Intact status of MLH1, c) Loss status of PMS2, d) Intact status of PMS2.}
\label{comet}
\end{figure}

The MMR status is a key predictive biomarker for colorectal cancer, essential for selecting patients for immunotherapy \cite{worthley2010colorectal}. Approximately 15\% of colorectal cancer patients exhibit microsatellite instability (MSI) \cite{luchini2019esmo}, and clinical guidelines recommend MSI testing for all patients diagnosed with colorectal cancer \cite{nice}. MMR status is typically assessed using IHC staining of MutL homologue 1 (MLH1), postmeiotic segregation increased 2 (PMS2), mutS homologue 2 (MSH2) and mutS homologue 6 (MSH6) proteins, with expression quantified as either positive/intact or negative/loss  \cite{awan2022deep}. Fig. \ref{comet} contains examples of ROIs from slides with differing MLH1 and PMS2 status.

The main contributions of this study are, therefore, as listed below:
\begin{itemize}
  \item We introduce a novel dual hierarchical attention model for the classification of gigapixel WSIs and propose a dynamic joint loss function to train the model in an end-to-end.
  \item We also propose a novel sampling strategy to increase the chances of selecting informative and spatially distinct image tiles and have reduced the percentage of overlapping tiles by approximately 60\%.
  \item We demonstrate that the proposed method achieves performance better or comparable to state-of-the-art methods while analyzing less than 10\% of the WSI tissue regions at the highest magnification.
  \item We show that the proposed method significantly reduces the time required to process a slide during inference by more than 75\%. 
\end{itemize}

\section{Literature Review}
\label{Literature Review}


\subsection{ Methods for WSI Classification}

WSI classification methods can be broadly categorized into either patch exhaustive or patch selective methods \cite{zhang2021joint}. The patch exhaustive approach, which commonly employs the "sliding window" paradigm, assigns the slide-level label to every patch, in the absence of patch-level annotations \cite{sanyal2021dan, bashir2020automated, qaiser2019learning}. Reducing the costs associated with this approach is a key research focus in CPath. Patch selective methods including weakly supervised learning and attention-based methods  \cite{ilse2018attention,bilal2023aggregation} that aim to identify the most relevant regions or patches as opposed to random patch selection \cite{mukundan2017robust, wulczyn2020deep}.  Yang \emph{et al.} \cite{yang2019guided} used region-level annotations to guide the model to focus on diagnostically relevant areas for breast cancer classification.  Thandiackal \emph{et al.} \cite{thandiackal2022differentiable},  Liu \emph{et al.} \cite{liu2023dsca} and Karthik \emph{et al.} \cite{karthik2024hmarnet} proposed methods that aggregate features extracted across multiple magnifications for downstream tasks.  In a similar vein, Zhang \emph{et al.} \cite{zhang2021joint} used a thumbnail to generate attention distributions at different magnifications. However, generating multiple attention maps at different magnifications is computationally expensive, therefore the proposed approach employs only a single attention map. Kong \emph{et al.} \cite{kong2021efficient} proposed using attention maps to "zoom" into informative slide regions, but their method wasn't evaluated at the whole slide level. Similarly, Katharopoulos \emph{et al.} \cite{katharopoulos2019processing} used low resolution images for  generating attention distributions and high-resolution image tiles for classification, but their model was limited to H\&E slides. Our proposed method uses soft attention at the slide-level and further employs hard attention at the tile-level, employs a custom sampling method for tile selection and was evaluated on IHC-stained WSIs.

The application of deep reinforcement learning (RL) techniques in the fields of medical imaging and computational pathology is a subject of considerable ongoing research \cite{hu2023reinforcement,bentaieb2018predicting, zhou2021deep, hering2020multiple, maksoud2020sos}. Barata \emph{et al.} \cite{barata2023reinforcement} introduced a RL model for skin cancer diagnosis by using an expert-generated reward table to guide the model.  Zheng \emph{et al.} \cite{zheng2023learning} employed a SeAgent to identify ROIs and a DeAgent to perform more detailed detection for melanoma histopathology images. However, their method required the use of an independently trained tumor classification model to serve as a guide for the SeAgent. Xu \emph{et al.} \cite{xu2019attention} proposed a hybrid attention model where a soft attention mechanism was applied to patches extracted using hard attention. In contrast, our proposed approach employs soft attention as the initial step, assigning continuous scores to image areas to create a  heatmap. This helps in understanding the relative importance of different regions for classification. Moreover, while their model classified images using 15\% of the image, our method uses less than 10\% of the tissue area in a slide.

\begin{figure*}[h]
\centering
\includegraphics[width=\textwidth]{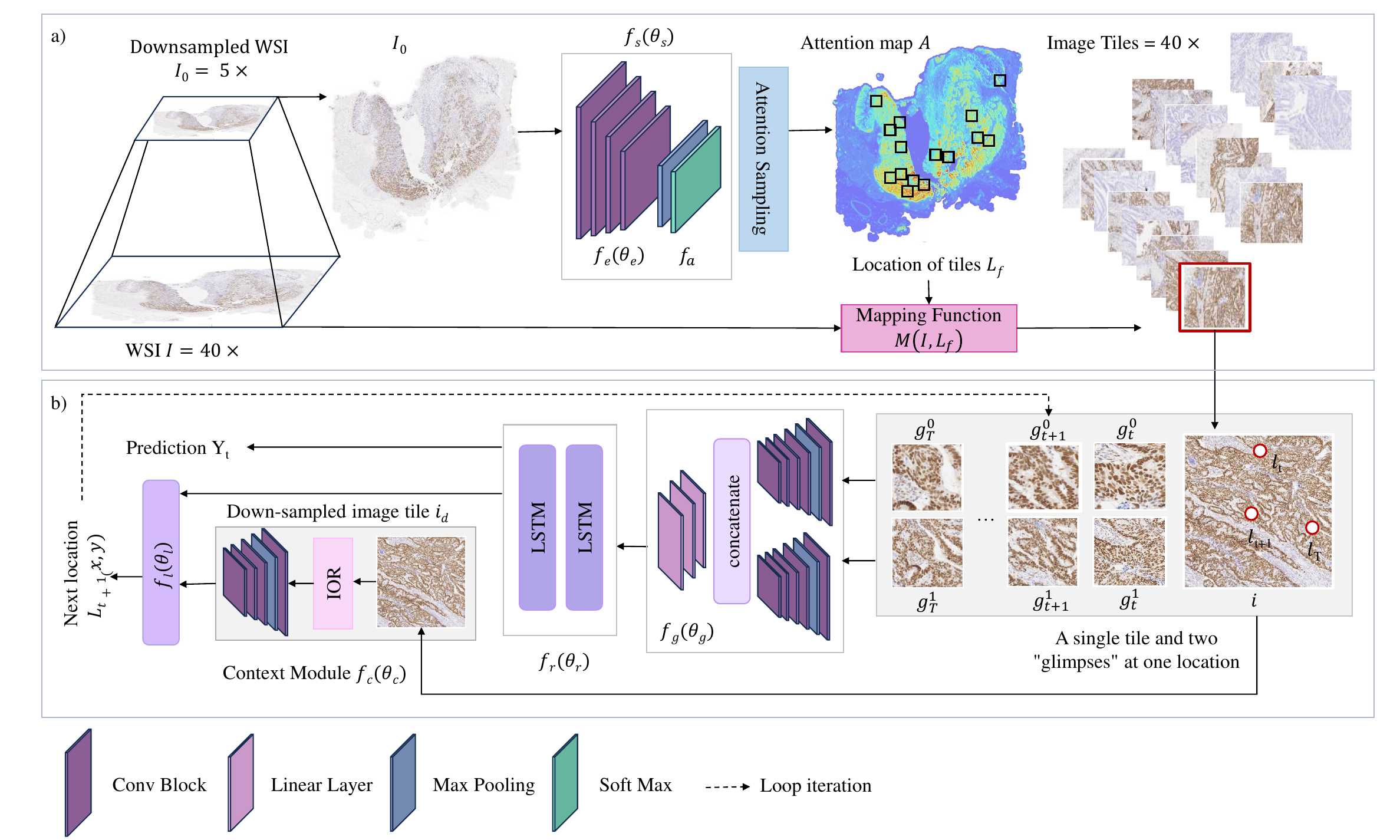}
\caption{An overall concept of the proposed dual attention method; a) Depicts the downsampled WSIs $I_{0} \in R ^{ h \times w \times C}$ passed as input to the soft attention module which produces the attention maps. The attention sampling method refines the attention maps and extracts image tiles of size $2048 \times 2048$ at $40 \times$ from the locations $L_{F}$ depicted in the diagram; b) Each extracted image tile is passed to the hard attention module which extracts $T$ multi resolution glimpses from the image tile at 20$\times$ and 40$\times$. Information from all the glimpses is processed to come to assign a score to the image tile. Additionally, a downsampled version of the image tile is used to provide contextual information in selecting the location of the glimpses. }
\label{fig3}
\end{figure*}

\subsection{IHC Scoring}

While the ``gold standard" for IHC scoring involves visual examination by expert pathologists, several methods have employed deep learning for IHC scoring  \cite{ayad2015comparative,rodner2017deep, mukundan2019analysis,saha2018her2net,pitkaaho2016classifying, kabir2024utility}.  Pham \emph{et al.} \cite{pham2023interpretable} developed a HER2 slide scoring pipeline in which pathologists identify ROIs to train a tumor segmentation model for patch extraction. These patches were then used to train the model in a weakly supervised manner. Selcuk \emph{et al.} \cite{selcuk2024automated} utilized pyramid sampling at multiple spatial scales to score IHC stained breast cancer bright-field images. Jha \emph{et al.}\cite{jha2024development} employed a two-step approach wherein they first performed membrane and nuclei segmentation and subsequently computed morphological features for the WSIs which were used to predict the HER2 scores. Qaiser \emph{et al.} \cite{qaiser2019learning} has previously shown that RL works well for IHC images. Their model employed DRL to effectively identify patches by way of a parametrized policy that prevented the model from revisiting locations it had already analyzed. However, the selection of these patches takes place in the context of a single image tile. This necessitated the identification of diaminobenzidine (DAB) regions and consequently the extraction of image tiles using the sliding window paradigm. The proposed model aims to automate the tiling process by identifying ROIs using soft attention mechanisms and applying reinforcement learning as a subsequent step only on the selected tiles, thereby presenting a cohesive end-to-end alternative and overcoming the challenges inherent in patch-based models. The proposed method therefore is able to classify IHC-stained WSIs by analyzing less than 10\% of the tissue area in a slide.

\section{The Dual Attention Method}
In routine clinical practice, when examining a tissue slide, a pathologist typically begins by viewing the different tissue components at a lower magnification to identify diagnostically relevant high-power fields (HPFs), and then proceeds to examine the morphological details of these selected HPFs at a higher magnification. We draw inspiration from a pathologist's visual assessment of a tissue slide and propose a method consisting of two main components which can function simultaneously.  The first component is a soft attention model which takes as input a high-level (low magnification) view of the WSI to determine various ROIs,  as illustrated in Fig.\ref{fig3}.a. We extract image tiles or HPFs (at higher magnification) from these high attention ROIs for classification using a custom sampling method that ensures each extracted tile is informative and spatially distinct.. The second component is the hard attention classification model, which further extracts a sequence of multi-resolution glimpses or patches from each image tile, as illustrated in Fig.\ref{fig3}.b.  The location of each glimpse is decided upon by the preceding glimpses and information from all the glimpses is collectively processed to come to a classification decision for the HPF in question. This process is repeated iteratively until all image tiles from the slide are processed,  allowing for the assignment of a slide-level score to the WSI.  We have designed the entire workflow in an end-to-end manner and proposed a joint loss function that simultaneously trains both the soft and hard attention components. The illustration of our proposed workflow is shown in Fig. \ref{fig3}. In the following sections, we explain both the soft and hard attention modules and their respective loss functions as well as the attention sampling procedure and the joint loss function.

\begin{figure*}[!t]
\centerline{\includegraphics[width=\textwidth]{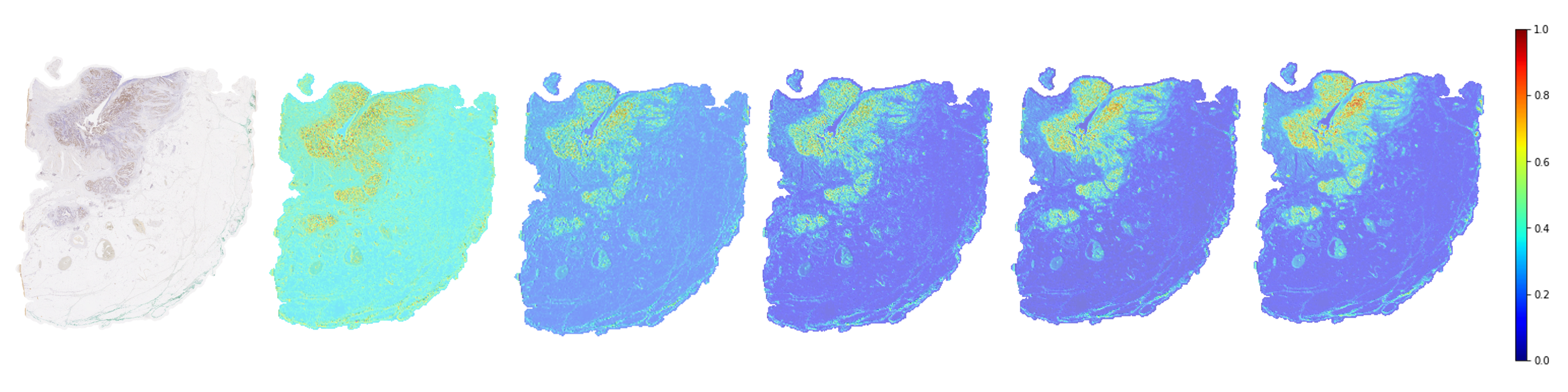}}
\caption{(left to right) The first column shows the original MLH1 WSI. The remaining columns show the evolution of the attention maps along different epochs as training progresses for the given slide.}
\label{fig4}
\end{figure*}

\subsection{Soft Attention Model}
\label{soft attention model}
The Soft Attention Model (SAM) $f_{s}{(\theta_{s})} $ with learnable parameters $\theta_{s}$, aims to discover potential ROIs in the input image that could prove conducive to the classification task at hand\cite{katharopoulos2019processing}. Let the WSI be denoted as $I \in R ^{ H \times W \times C}$ where $H, W$ and $C$ represent the height, width, and channels of the image respectively. 
We initially take as input a down sampled version of the WSI scaled by a factor $s$, $I_{0} \in R ^{ h \times w \times C}$ where $h \ll H$ and $w \ll W$. Let $f_{e}(.)$ parameterized by $\theta_{e}$, represent a function of feature extractor network consisting of a convolutional neural network (CNN) with rectified linear unit (ReLU) non-linearities that extracts discriminative features from the input image $I_{0}$. Employing an attention mechanism $f_{a}(.)$ on the extracted features is similar to learning a probability distribution over the pixels of the image such that the probability always sums up to one as shown in eq. \eqref{eq1}.
\begin{equation}
\sum_{j=0}^{N} f_{a}(I_{0},\theta_{a})_{j} = 1\label{eq1}
\end{equation}
where $I_{0}$ denotes a downsampled image passed to the model and $N = h \times w$ denotes the number of pixels. The probabilities correspond to the relevance of the areas for the classification task. The overall definition of the proposed soft attention model is given in eq. \eqref{eq2}. 
This results in an attention map $A \in R ^{h \times w}$ which highlights potentially informative regions in the image $I_{0}$. 
\begin{equation}f_{s}(I_{0},\theta_{s}) = f_{a}( f_{e}(I_{0},\theta_{e}),\theta_{a}) \label{eq2}\end{equation}

\subsection{Attention Sampling}
The attention map $A$ contains both high and low attention regions, where the attention probabilities indicate the relative importance of different regions within the image for the classification task. Fig. \ref{fig4} shows attention maps for a single slide as training progresses. We sample a set of locations from the high attention regions without replacement in order to extract a set of representative image tiles from the WSI, as depicted in Fig.\ref{fig3}.a. 

A major challenge in this regard, is the selection and consequent extraction of a set of informative and spatially distinct image tiles. In order to ensure reliable classification outcomes, the image tiles must encompass important information relevant to the classification task and they must be extracted from different regions within the WSI to capture a comprehensive view of the underlying tissue before assigning the slide a final label. To achieve our goal of extracting a set of spatially diverse image tiles we adopted a two step attention sampling procedure. The proposed sampling method introduces controlled noise into the attention distribution prior to the sampling procedure. Additionally, during sampling we employ a distance-based strategy to select image titles that are maximally separated.

Our custom sampling method injects noise only within the tissue regions to prevent the selection of image tiles from the irrelevant background regions. To this effect, we created a noise matrix, $n \in R ^{h \times w}$, by generating random numbers sampled from a uniform distribution in the range $[low, high)$ where both $low$ and $high \in R$ and $high > low$. To extract the image tiles, we first normalize the attention distribution $A$ to the range $[0,1)$, denoted as $\bar{A}$, using min-max normalization. In order to focus the “noise” on the relevant areas within the slide we multiplied the noise vector with the masks $m$, created by thresholding the attention maps, explained further in Section \ref{experimental setup}. Additionally, we also multiplied the normalized attention map with the mask which gives us our final attention distribution.  

\begin{equation}F_{a} = (m\cdot n) +  (m\cdot \bar{A})\end{equation}

Thereafter we performed top-$k$ sampling on our final attention maps with the target noise. This gives us a set of location indices. We perform an additional distance-constrained point selection step (DCPS) as part of our sampling procedure, whereby potential locations are narrowed down based on a distance metric, to arrive at the final set of sampling locations and corresponding image tiles. If we denote $N$ to represent the desired number of tiles to be extracted from the WSI then for each image $I_{0}$ we initially sampled $2 \times N$ locations $L \in R ^{N \times 2}$ from the high probability regions. Out of this larger pool of locations, some may be spatially close to each other thereby contributing to data redundancy. To reduce the likelihood of obtaining overlapping image tiles, we select tiles that are maximally separated. To achieve this, we initially created an empty set to represent our “shortlisted” final locations,  $L_{F} \in R ^{N}$. At each time-step $t$, a location index $L_{t}= (x ,y)$ is only added to $L_{F}$ if the euclidian distance, $d$, between the coordinates $L_{t}$ and all coordinate pairs previously in $L_{F}$ is greater than a certain threshold, $d_{t}$. 



We define a mapping function $M(I, L_{F})$ that maps each coordinate pair in $L_{F}= (x, y)$ to a location in the WSI $I \in R ^{ H \times W \times C} (X, Y)$ using the pre-defined scaling factor $s$  between the WSI $I \in R ^ { H \times W \times C}$ and image $I_{0} \in R ^{h \times w \times C}$ \cite{eriksson_hu_2018}.


\subsection{Soft Attention Regularization}
The soft attention mechanism represents a multinomial distribution over the pixels in the image. Let $E$ be the multinomial entropy function for the attention maps. To ensure the attention distribution does not focus on a small number of locations where it receives maximum reward, we used an entropy regularizer \cite{katharopoulos2019processing} \cite{zhang2021joint}. This approach attempts to resolve the exploration-exploitation dilemma by preventing the model from quickly assigning high probabilities to a few selected locations. Instead, it encourages the model to ‘explore’ various locations:  
\begin{equation} L_{SA} = \beta E(f_{s}(I_{0}; \theta_{s})) \label{eq3}\end{equation}
where $\beta$ denotes the entropy regularizer and $f_{s}$ denotes the soft attention mechanism. The above formulation results in a relatively uniform attention distribution due to the usage of entropy. 

\begin{figure*}[!t]
\centerline{\includegraphics[width=\textwidth]{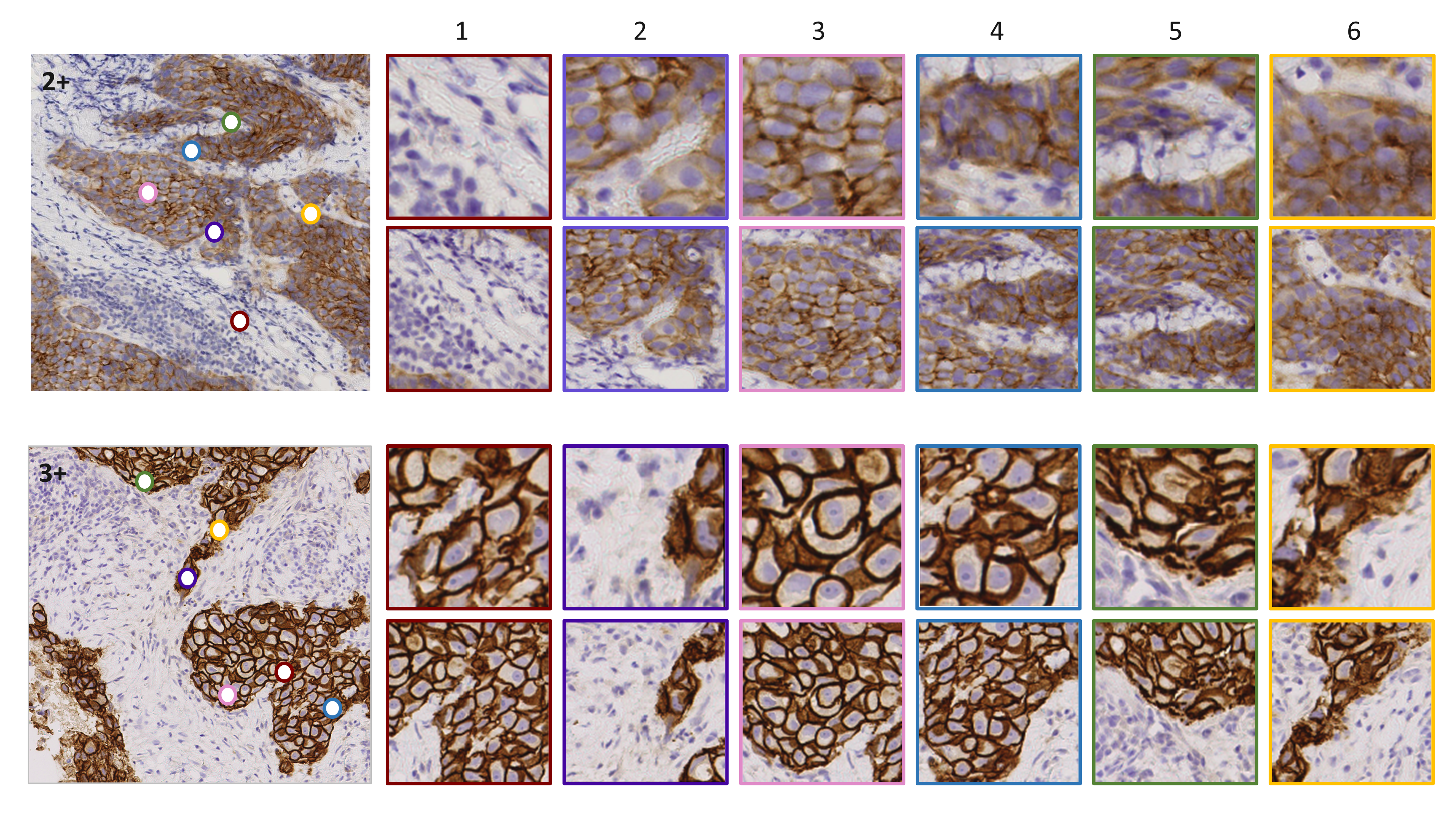}}
\caption{Examples of image tiles of HER2 scores 2+ (top row) and 3+ (bottom row). The colored circles represent the locations of the six glimpses used for the prediction of the HER2 score for the respective tiles. The remaining images show the glimpses extracted at the circled locations at 40$\times$ and 20$\times$.}
\label{fig5}
\end{figure*}

\subsection{Hard Attention RL Model } \label{Hard Attention RL Model}
The tiles extracted from high attention regions are still large enough to contain irrelevant areas. Moreover training a CNN directly with the tiles remains infeasible due to computational and memory constraints. 
Therefore, we proposed a hierarchical dual attention method that incorporates soft attention at the WSI-level to extract relevant tiles and then hard attention at the tile-level for the underlying task. Our soft attention module provides us with a set of $N$ image tiles of size $2048 \times 2048$ at resolution 40$\times$ extracted from the WSI as depicted in Fig.\ref{fig3}.a. Thereafter, we employ a retina-inspired hard attention mechanism in our classification model to selectively analyze informative parts of a single image tile as illustrated in Fig.\ref{fig3}.b. 

Our model can be formalized by what is known as a Partially Observable Markov Decision Process (POMDP) in reinforcement learning literature. The image tile can be thought of as an environment and the neural networks collectively act as an agent, a decision-making entity, that linearly interacts with the environment.  

Considering a single image tile $i$ sampled from $I$, we extract multi-resolution patches from the tile to reach a classification decision. At each time-step $t$, the model receives a state from the environment, consisting of two multi-resolution glimpses centered around an individual location $l_{t} = (x_{t}, y_{t})$. The two glimpses $g_{t} = ( g_{0t} , g_{1t} ) \in R^{hg {\times} wg {\times} C}$ of size $128 \times 128$ are extracted at resolutions 40$\times$ and 20$\times$ respectively, mimicking the peripheral vision around a single point of focus. A CNN $f_{g}$ parameterized by $\theta_{g}$, acts as a non-linear function which takes as input these two images and aggregates the features extracted from these two glimpses to create a feature representation vector $\textbf{v}_{gt}$. To capture both the semantic and spatial information, the features of the location coordinates $\textbf{v}_{lt}$ are also extracted by way of a linear layer and combined with the feature vector $\textbf{v}_{gt}$ to obtain $\textbf{v}_{t} = \textbf{v}_{gt} \times \textbf{v}_{lt}$. 
We use multiple such glimpses analogous to a sequence of saccades. An example of this sequence of glimpses can be found in Fig. \ref{fig5}.


The backbone of the hard attention classification model is a recurrent neural network (RNN). In the absence of unrestricted access to the entire image tile $i$, the RNN sequentially builds up an internal representation of the environment using these glimpses. To retain information from earlier time-steps and to learn the spatial dependencies between the glimpses, we use a long short-term memory (LSTM) based RNN. At each timestep $t$ the RNN $f_{r}({\theta_{r}})$ with learnable parameters $\theta_{r}$ processes the aggregated features of the glimpses to update the parameters of its internal hidden states (memory units), $h_{t}$.  
\begin{equation} h_{t+1} = f_{r}(h_{t}, \textbf{v}_{t} ; \theta_{r}) .\label{eq4}\end{equation}

Given the current state or “glimpse” of the environment, the CNN and RNN collectively act as an agent to predict the location of the next glimpse $l_{t+1}= (x,y)$ and a classification decision. Another CNN $f_{c}$ was employed to embed contextual awareness in the model and to aid in the selection of the ROIs. This model with learnable parameters $\theta_{c}$ takes as input a down-scaled version of the original image tile $i_{d}$, to provide contextual information and assist in the selection of locations for the glimpses.  

To ensure that the model selects spatially distinct locations an “Inhibition of Return” (IOR), which suppresses the textural information of previously visited locations by blacking out those regions, was incorporated into the process. The outputs of the last hidden layer of the recurrent neural network and the context module are combined using the Hadamard product formula to obtain a feature vector. The location network $f_{l}(\theta_{l})$ linearly transforms the resulting feature vector to predict the next location $l_{t+1}= (x_{t+1},y_{t+1})$. 

In order to model human visual attention, the entire process is repeated iteratively for $T$ time-steps on a single image tile. At the end of this sequence of saccades $(g_{t-1}, g_{t}, g_{t+1}...g_{T})$, referred to as an "episode" in RL literature, the models predict the final classification label $Y_{T}$ for the image tile $i$ under consideration.

\subsection{Hard Attention Loss function}
All the individual components of the retina-inspired hard attention model were trained in an end-to-end fashion, where we optimized the loss obtained from each neural network present in the overall framework i.e. the glimpse network $f_{g}$, the recurrent model $f_{r}$, the contextual CNN $f_{c}$ and the location network $f_{l}$. In order to optimize the loss, the hard attention model learns a parameterized DRL policy $\pi$ where each state is mapped to a set of actions by maximizing the sum of the expected reward of said actions while following the parameterized policy.  In our research study, the actions consist of coordinates of the next location $l_{t+1} (x_{t+1},y_{t+1})$ from where the multi-resolution glimpses will be extracted and classification decision $Y_{T}$, provided the input is the glimpse $g_{t} = ( g_{0t} , g_{1t} )$ at time-step $t$ from location $l_{t}(x_{t},y_{t})$ and the down sampled version of the image tile $i_{d}$.
\begin{equation} \pi((l_{t+1}(x_{t+1},y_{t+1}),Y_{T} )|(\textbf{v}_{t},i_{d});\theta_{HA}) .\label{eq5}\end{equation}
where $\theta_{HA} = \theta_{g},\theta_{r},\theta_{c},\theta_{l}$. The reward is calculated using the following equation on all glimpses for a given tile. 
\begin{equation} r_{t}=\left\{
  \begin{array}{@{}ll@{}}
    1  \text{ if}\ GT=Y_{t} \\
    0  \text{ otherwise}
  \end{array}\right.\label{eq6}\end{equation}
where $r_{t}$ and $Y_{t}$ denote the reward and predicted score for each glimpse respectively and GT denotes the ground truth. 

The final reward is calculated by taking sum of all rewards of the entire glimpse episode $(g_{t-1}, g_{t}, g_{t+1}...g_{T})$, where 1 is a weighing factor (a concept commonly used in RL literature) as shown in eq. \ref{eq}.

\begin{equation}\sum_{t=0}^{T} 1_{t-1} r_{t}\label{eq}\end{equation}

The parameterized policy $\pi$ is optimized with the help of maximizing the model's performance with respect to the reward and this objective can be achieved by the reverse of gradient descent. Rather, to maximize the policy gradients the REINFORCE rule has been employed where in each episodic scenario the actions leading to higher rewards are computed using the gradients and log probabilities such that the actions with low rewards are minimized. In order to reduce the intra-class variance, a common drawback of the REINFORCE method, a baseline function was introduced by normalizing the model parameters. 
\begin{equation}\triangledown L_{\theta} = \sum_{t=0}^{T} \triangledown _{\theta} log \pi((l_{t+1},Y_{T} )|(\textbf{v}_{t},i_{d});\theta_{HA}) (R_{t}- B_{t})
\label{eq7}\end{equation}  

To prevent the model from extracting glimpses from the same location, the model was discouraged from visiting previously attended locations. This is achieved by essentially "blacking out" regions from where the previous glimpses have been extracted within the down sampled image tile $i_d$ which provides contextual information to the model. An additional term was introduced where the sum of the overlapping bounding boxes for the glimpse the locations was calculated. This term was incorporated into the final loss function to encourage the model to select spatially distinct locations.
\begin{equation}L_{bb} =\frac{1}{_{T}C_{2}} \sum_{t=0}^{T} \sum_{tb=t+1}^{T} b(g_{t})\cap b(g_{tb}) \label{eq8}\end{equation}
where $_{T}C_{2}$ is the number of possible combinations of T distinct objects taken 2 at a time and $b$ denotes a function that calculates the bounding box around glimpse $g$ at time-step $t$.

Moreover, to penalize wrong predictions in accordance with their clinical significance, an additional task-specific regularization term was added, as shown below:
\begin{equation}L_{s} = |Y_{t} - GT|\label{eq:ls}\end{equation}

$L_{s}$ is a measure of the absolute difference between the ground truth values and the predicted values, considering both are numerical scores. Therefore, this particular component in the loss function will penalize the model based on how “far apart” the ground truth and predicted values are. 

Finally, these losses are trained as in eq. \eqref{eq:Lha}, where $\delta$ is used to control the effect of $L_{s}$ and $L_{bb}$.
\begin{equation}\label{eq:Lha} L_{HA} = L_{\theta} + \delta (L_{bb} + L_{s}) \end{equation}

\subsection{Joint Loss Function}

We have proposed a method of training both components in an end-to-end manner. Since the retina-inspired classification model takes much longer to train than the soft attention model, the loss of the soft attention model is incorporated into the final loss function using a decaying coefficient. This allows us to train the model end-to-end with a dynamic loss function.

The joint loss function is as in eq. \eqref{eq:Ljq}, where, $\alpha$ denotes the loss-weightage hyper-parameter and $e$ denotes the epoch number. 
\begin{equation}\label{eq:Ljq} L_{J} = L_{HA}+ (\alpha ^{e} \times L_{SA})\end{equation}

\section{Datasets}
We have evaluated the performance of the proposed method on two separate use-cases to showcase the the generalizability of the approach, the prediction of the Loss/Intact Status of two MMR biomarkers, MLH1 and PMS2 in colorectal cancer and HER2 prediction for invasive breast carcinoma. 

\subsection{COMET MMR Dataset}
We evaluated the performance of the proposed model on an internal dataset, the COMET-MMR dataset, which comprises 72 WSIs obtained from patients with colorectal cancer. The dataset was collected using the East Midlands Research Ethics Committee (reference 11/WM/0170). The slides were scanned using the Omnyx VL120 scanner and stained with 4 MMR markers. These proteins form pairs, namely MLH1 with PMS2 and MSH2 with MSH6 where a loss in expression in MLH1 and MSH2 results in a loss of expression in their respective counterparts but not the other way around \cite{li2008mechanisms}.  Routinely, MMR status is assessed by performing IHC staining for all four proteins. The expression in MMR biomarkers is quantified as either positive/intact if the DAB stain detected in the cancerous nuclei is of similar or stronger intensity as in the normal tissue cells or negative/loss if weak staining is detected in the tumor nuclei in comparison to normal tissue cells \cite{awan2022deep}. 
For this study, we only considered MLH1 and PMS2. The dataset consisted of 14 negative (loss) cases and 58 positive (intact) cases for both MLH1 and PMS2. For our experiments, we performed five-folds cross-validation to create training, validation and testing datasets due to the imbalanced dataset. The ground truth (GT) for the WSI-level MMR scores were provided by an expert pathologist \cite{awan2022deepmsi}.  

\subsection{HER2 Dataset}
The HER2 challenge dataset  \cite{qaiser2018her} is publicly available, covered by the Nottingham Research Ethics Committee 2 approval no. REC 2020313 (R\&D reference 03HI01). The contest dataset comprised a total number of 172 WSIs obtained from 86 invasive breast cancer cases. Each case contains an IHC stained HER2 slide and the corresponding H\&E slide. Similar to routine clinical practice, for our experiments, we only take the IHC-stained slides into consideration for assigning HER2 scores. The training dataset consisted of 52 cases, with 13 cases belonging to each of the four scores, while 28 cases were reserved as part of the test dataset. On average the images were made up of approximately $10^{10}$ pixels and were scanned using the Hamamatsu NanoZoomer C9600. The WSIs possessed a multi-resolution pyramidal structure in the range 4$\times$ to 40$\times$. The ground truth (GT) for each case was obtained from clinical reports marked by at least two histopathologists and included the slide-level HER2 score for each case $(0-3+)$.  The slide is assigned a negative score (0,1+) if weak or incomplete membrane staining is observed in no more than 10\% of the invasive tumor cells. Tissue slides with non-uniform staining are considered borderline or weakly-positive and assigned a score of 2+. Such cases are recommended for further fluorescence in-situ hybridisation (FISH) testing. Tissue slides are scored HER2+ and assigned a score of 3+ if strong membrane staining is observed in more than 10\% of the invasive tumor cells \cite{qaiser2018her}.  
For our experiments, we performed four-folds cross-validation to compare with \cite{qaiser2019learning}  and have used the 28 slides in the contest testing dataset as the test data \cite{qaiser2018her}.

\section{Experiments}
\subsection{Experimental Setup}
\label{experimental setup}

\begin{table}[ht]
\centering
\scriptsize
\caption{Detailed architecture specifications for the main components of the soft attention model. The output shapes represent (batch size, channels, height, width). } 
\label{table:sa_architecture}
\begin{tabularx}{\textwidth}{|X|X|X|X|}
\hline
\textbf{Layer Group} & \textbf{Layers} & \textbf{Output Shape} & \textbf{Input} \\ \hline
\multicolumn{4}{|c|}{\textbf{Attention Model}} \\ \hline   
\textbf{Input Layer} & Image & (1, 3, $h$, $w$) & \\ \hline
\textbf{Convolution Block 1}     & Conv2D (8, 3$\times$3), ReLU & (1, 8, $h-2$, $w-2$) & Input Layer  \\ \hline
\textbf{Convolution Block 2}     &  Conv2D (16, 3$\times$3), ReLU & (1, 16, $h-4$, $w-4$)     &  Convolution Block 1    \\ \hline
\textbf{Convolution Block 3}     & Conv2D (32, 3$\times$3), ReLU & (1, 32, $h-6$, $w-6$)       & Convolution Block 2   \\ \hline
\textbf{Convolution Block 4}     & Conv2D (1, 3$\times$3)       & (1, 1, $h-8$, $w-8$)       &  Convolution Block 3  \\ \hline
\textbf{Attention Layer} & MaxPool2d (8$\times$8), SoftMax & (1, 1, floor(($h-8)/8$), floor(($w-8)/8$)) & Convolution Block 4 \\ \hline
\end{tabularx}
\end{table}

\begin{table}
\centering
\scriptsize
\caption{Detailed architecture specifications for the main components of the hard attention model. The output shapes of the input layer represent (number of image tiles, height, width, channels) whereas the output shape of glimpse extraction layer represents (number of image tiles, number of glimpses, channels, height, width). } 
\label{table:RL_architecture}
\begin{xltabular}{\textwidth}{|X|X|X|X|}
\hline \multicolumn{1}{|c|}{\textbf{Layer Group}} & \multicolumn{1}{c|}{\textbf{Layers}} & \multicolumn{1}{c|}{\textbf{Output Shape}}  & \multicolumn{1}{c|}{\textbf{Input}} \\ \hline 
\endfirsthead

\multicolumn{4}{|c|}%
{\tablename\ \thetable{} -- continued from previous page} \\
\hline \multicolumn{1}{|c|}{\textbf{Layer Group}} & \multicolumn{1}{c|}{\textbf{Layers}} & \multicolumn{1}{c|}{\textbf{Output Shape}}  & \multicolumn{1}{c|}{\textbf{Input}} \\ \hline 
\endhead

\hline \multicolumn{4}{|r|}{{Continued on next page}} \\ \hline
\endfoot

\hline
\endlastfoot

\multicolumn{4}{|c|}{\textbf{Glimpse Network}} \\ \hline   
\textbf{Input Layer}      & 20$\times$ image , location, 40$\times$ image             & (1, 128, 128, 3)      
                                                   (1, 2) (1, 2048, 2048, 3)  & \\ \hline   
\textbf{Glimpse Extraction}   &           & (1, 2, 3, 128, 128) & Input Layer \\ \hline                  
\textbf{Convolution Block 1}     & Conv2D (32, 3$\times$3), ELU, MaxPooling        & (1, 2, 32, 63, 63) & Glimpse Extraction Layer  \\ \hline
\textbf{Convolution Block 2}     &  [Conv2D (64, 3$\times$3), ELU]$\times$2                  & (1, 2, 64, 15, 15)     &  Convolution Block 1    \\ \hline
\textbf{Convolution Block 3}     & Conv2D (128, 3$\times$3), ELU, Dropout         & (1, 2, 128, 7, 7)       & Convolution Block 2   \\ \hline
\textbf{Convolution Block 4}     & Conv2D (128, 7$\times$7), ELU, Dropout         & (1, 2, 128, 1, 1)       &  Convolution Block 3  \\ \hline
\textbf{Flatten Layer}     & Flatten & (1, 2, 128)     & Convolution Block 4 \\ \hline
\textbf{Concatenation}     & Concatenate  & (1, 256)    &  Flatten Layer          \\ \hline
\textbf{Fully Connected 1} & Linear Layer, ReLU                   & (1, 384)        &   Concatenation     \\ \hline
\textbf{Fully Connected 2} & Linear Layer, ReLU                   & (1, 256)         &   Fully Connected 1    \\ \hline
\textbf{Location Layer}   & Linear Layer, ReLU, Linear Layer      & (1, 256)    & location \\ \hline

\textbf{Output Layer} & Addition, ReLU         & (1, 256)  & Fully Connected 2, Location Layer   \\ \hline
\multicolumn{4}{|c|}{\textbf{Context Network}} \\ \hline   
\textbf{Input Layer}      & image tile , location        & (1, 128, 128, 3) (1, 2) & \\ \hline
\textbf{Convolution Block 1} & Conv2D (32, 3$\times$3), Tanh, MaxPooling & (1, 32, 63, 63) & Input Layer  \\ \hline
\textbf{Convolution Block 2} & Conv2D (64, 3$\times$3), Tanh            & (1, 64, 31, 31)  &   Convolution Block 1 \\ \hline
\textbf{Convolution Block 3} & Conv2D (128, 3$\times$3), Tanh            & (1, 128, 15, 15)  & Convolution Block 2  \\ \hline
\textbf{Convolution Block 4} & Conv2D (128, 3$\times$3), Tanh , Dropout  & (1, 128, 7, 7)  &  Convolution Block 3   \\ \hline
\textbf{Final Convolution}  & Conv2D (256, 7$\times$7), Tanh           & (1, 256, 1, 1)  & Convolution Block 4    \\ \hline
\textbf{Output Layer}      & Flatten  & (1, 256)      &  Final Convolution     \\\hline
\end{xltabular}
\end{table}

The WSIs were down-sampled by a factor of 32 to level 5 at 1.25$\times$ as $I_{0}$. To create the mask, we first employed Otsu’s method to determine a threshold between the background region and the tissue section followed by morphological operations to further refine the mask structure and remove the control tissue \cite{pocock2021tiatoolbox} \cite{wrap89986}. The soft attention model consisted of a CNN with ReLU activation functions and max-pooling layer and softmax layers where we increased channels in subsequent layers \cite{katharopoulos2019processing}. The architectural specifications for the key components of the soft attention model are shown in Table. \ref{table:sa_architecture}. For simplicity, we have assumed a batch size of 1 in the table. Furthermore, since the soft attention model processes a down sampled version of the original WSI, the height and width of the image can vary. Therefore, we use the same notation introduced in Section \ref{soft attention model} and represent the height and width of the down-sampled image as $h$ and $w$ in the table.

The HER2 score for a slide is determined by the intensity of staining. Therefore, attention maps were used to identify regions within the image most likely to exhibit HER2 positivity, by focusing on the intensity of DAB staining, using adaptive thresholding. In contrast, in the COMET-MMR dataset, the binary label assigned depends on the presence of the stain marker rather than the intensity of staining. Thus, \emph{k}-means clustering with three clusters was used to categorize the tissue regions into high attention areas, low attention areas and background regions. The attention masks, $m$ were created by selecting regions identified as high-attention areas through these methods, and these masks were then used to select image tiles for downstream analysis. For HER2, we initially selected 20 locations and then we distilled $N = 10$ spatially distinct tiles by identifying the tiles with the maximum Euclidean distance. Similarly, for the COMET-MMR dataset, $30$ locations were pooled to shortlist $N = 15$ spatially distinct tiles as $L_{F}$. Using empirical evaluation we found that $N = 10$ is best trade-off for HER2 and $N = 15$ is the best trade-off for the COMET dataset.

Each of the coordinate pairs $L_{F}$ was mapped to a location in the original WSI using a scale factor $s = 32$. The image tiles {$i_0 \cdots i_N$} of size $2048 \times 2048 \times 3$ were extracted at resolution $40\times$. Furthermore, we performed standard data augmentations: random rotations $(0, 90, 180, 270)$ and horizontal and vertical flipping on the extracted image tiles. 

During training, in each mini-batch step, $N$ image tiles were extracted from each WSI, with one WSI sampled from each class, resulting in an equal representation of all scores in a single batch. Given a single image tile $i$ from the batch, we extract six sequential multi-resolution glimpses of size $128 \times 128 \times 3$ at magnification levels $20 \times$ and $40 \times$. The location of the first glimpse was selected randomly. The architectural specifications for the key components of the hard attention model are shown in Table. \ref{table:RL_architecture}.  For simplicity, we have assumed a $N=1$ in the table.  The combined representation of the glimpse’s image and location features were of size $1 \times 256$ as shown in the table. \ref{table:RL_architecture}. The hidden layers of the RNN (LSTM) contained $256$ and $128$ neurons.

The Adam (Kingma \& Ba) optimizer \cite{kingma2014adam} and the StepLR scheduler were used for training the model. For the soft attention module, the initial learning rate was set to $0.0001$ with a weight decay of $0.00001$, while for the hard attention module, it was set to $0.001$. The scheduler for the soft attention module used a gamma of $0.1$ and a step size of $10,000$, whereas for the hard attention module, gamma was set to $0.99$ with a step size of $200$. Both optimizers used ${\text{betas}} = (0.9, 0.999)$. These choices were made to allow the soft attention module to converge gradually and in a stable manner whereas the hard attention module was configured with a learning rate suitable for fast decision-making for classification. In our experiments, we set the value of the regularizer as $1$ and the value of $\alpha$ as $0.5$.

\subsection{Comparison with state-of-the-art methods}
For the HER2 dataset, performance was compared with methods that competed in the HER2 Challenge \cite{qaiser2018her}. We also compared our approach with state-of-the-art (SOTA) methods, including ZoomMil \cite{thandiackal2022differentiable} and Attention based MIL \cite{ilse2018attention}, both of which utilize weak labels. The maximum number of epochs was set to 100 and early stop was determined by monitoring performance on the validation set. We used the official implementations of these methods.

Additionally, we proposed an experiment, Random Sampling, by randomly extracting image tiles of size $2048 \times 2048$ at $40 \times$, without replacement from the tissue area of the WSI. The tiles were extracted by using the Gumbel-max trick \cite{adams2013gumbel} instead of soft attention and passed to the RL-based hard attention component explained in Section \ref{Hard Attention RL Model} to obtain a classification prediction. This experiment aims to highlight the effectiveness of the soft attention component of the proposed method and to exhibit the advantages of employing an attention mechanism over random tile extraction. To further compare the performance of our framework, we proposed a second method, referred to as the Sliding Window Method. For each WSI, after the application of a tissue mask, a DAB mask was created by using a threshold of $0.85$ on a grayscale image. After the application of both masks, non-overlapping patches of size $224 \times 224$ at magnification $20 \times$ (to incorporate more contextual information) and containing at least $35\%$ tissue were extracted using the sliding window approach. These patches were subsequently processed by a ResNet-18 \cite{he2016deep} aligning with conventional deep CPath workflows.  This experiment demonstrates our ability to achieve better or comparable performance to conventional deep learning based CPath methods which can be both computationally inefficient and time-consuming. 

\section{Results}
\subsection{Quantitative Results}
\begin{table*}[h]
\caption{MMR Status Prediction : The table below presents the results of different models on the COMET MMR Biomarkers, mutL homologue 1 (MLH1) and post-meiotic segregation increased 2 (PMS2). The columns represent (in order) the F1-Score, AUROC, Precision and the percentage of tissue area and ROI selection process employed by each method.}
    \centering
    {\scriptsize
    \begin{tabularx}{\textwidth}{|X|X|X|X|X|X|X|}
    \hline
        \rule{0pt}{10pt} Method & F1-Score & AUROC & Precision & $p$-value  & \%Tissue area $\downarrow$ & ROI Selection \\ \hline
        \multicolumn{7}{|c|}{MutL homologue 1 (MLH1)}\\ \hline
         Proposed Method & $\textbf{0.88} \pm \textbf{0.09}$ & $\textbf{0.92} \pm \textbf{0.09}$ & $\textbf{0.96} \pm \textbf{0.04}$ & $p < 1e^{-10}$ &\textbf{6.53\%} & Automated \\ \hline
        Random Sampling & $0.66 \pm 0.06$ & $0.67 \pm 0.12$ & $0.83 \pm 0.04$ & $p < 1e^{-1}$ & 6.53\% & Random \\ \hline
        Sliding Window \cite{he2016deep} & $0.72 \pm 0.07$ & $\textbf{0.92} \pm \textbf{0.04} $ & $0.81 \pm 0.04$ & $p < 10^{-7}$ & 38.14\% & Sliding Window \\ \hline
        ZOOMMIL \cite{thandiackal2022differentiable} &$0.75 \pm 0.12$ & $0.67 \pm 0.13 $ & $0.87 \pm 0.13 $ & $p < 1e^{-5}$ & 6.53\% & Automated \\ \hline
        Attention based MIL \cite{ilse2018attention} & $0.87 \pm 0.11$ & $ 0.77 \pm 0.39 $ & $0.91 \pm 0.08 $ & $p < 1e^{-10}$ & 100\% & Sliding Window \\ 
        \hline
        \multicolumn{7}{|c|}{Post-meiotic segregation increased 2 (PMS2)}\\ \hline
        Proposed Method & $\textbf{0.83} \pm \textbf{0.08} $ & $0.89 \pm 0.09$ & $\textbf{0.94} \pm \textbf{0.05}$ & $p < 1e^{-7}$ &\textbf{6.53\%} & Automated \\ \hline
        Random Sampling  & $0.62 \pm 0.11$ & $0.64 \pm 0.04$ & $0.86 \pm 0.07$ & $ p < 1e^{-1}$ & 6.53\% & Random \\ \hline
        Sliding Window \cite{he2016deep} & $0.71 \pm  0.08 $ & $\textbf{0.93} \pm \textbf{0.08}$ & $0.80 \pm 0.05$ & $p < 10^{-5}$ & 39.04\% & Sliding Window \\ \hline
        ZOOMMIL \cite{thandiackal2022differentiable} & $0.72 \pm 0.08$ & $0.52 \pm 0.12 $ & $0.82 \pm 0.12 $ & $p < 1e^{-5}$ & 6.53\% & Automated \\ \hline
        Attention based MIL \cite{ilse2018attention} & $0.79 \pm 0.13$ & $0.87 \pm 0.14 $ & $0.86 \pm 0.09 $ & $p < 1e^{-7}$ & 100\% & Sliding Window \\ \hline
    \end{tabularx}
    }
    \label{table:two}
\end{table*}

\subsubsection{Comparison of Results on the COMET MMR Dataset}
In this section, we compare the proposed method with other approaches, presenting F1-scores, area under the receiver operating characteristic (AUROC) scores and Precision scores along with the $p$-values, percentage of average tissue area analyzed by each method and details of the ROI selection process, as illustrated in Table \ref{table:two}.  For the proposed model, ZoomMil and the Random Sampling baseline, we calculated the average probabilities of the $N = 15$ tiles selected per WSI to calculate the F1-scores and AUROC scores. For the Sliding Window baseline, for each slide, the top 15 probability values were used. Five-folds cross validation was performed with the same folds used in all the experiments, to ensure a fair comparison. 

As shown in Table \ref{table:two} the proposed method achieved the highest F1-scores and precision scores of ($0.88 \pm 0.09$ and $0.96 \pm 0.04$) and ($0.83 \pm 0.08 $ and $0.94\pm 0.05$) across both biomarkers MLH1 and PMS2 respectively. Additionally, it achieved a comparable AUROC of $0.92 \pm 0.09$, compared to the Sliding Window Method, which attained an AUROC of $0.92 \pm 0.04 $ for MLH1. This performance was attained by analyzing only $\approx7\%$ of the tissue area, whereas other methods achieved comparable performance while analyzing either the entire tissue area ($100\%$) or extracting tiles from the entire DAB region ($\approx39\%$), typically by using the sliding window approach.

To assess the statistical significance of the models' performance, we also conducted a one-sample t-test using the binary correctness of each model's predictions, where 1 indicates correct predictions and 0 indicates incorrect predictions. We aggregated the binary correctness data from all folds and conducted a single t-test on the combined test sets. The null hypothesis (\textbf{H$_0$}) assumes that the model's predictions are no better than random guessing, \textbf{H$_0 : \mu = 0.5$}. The alternative hypothesis (\textbf{H$_1$}) assumes that the models' prediction accuracy differs significantly from random guessing as \textbf{H$_1: \mu \neq 0.5$}. The proposed method demonstrated highly significant results for both MLH1 and PMS2, with $p$-values less than $1e^{-10}$ for MLH1 and less than $1e^{-7}$ for PMS2, indicating that the model performs significantly better than random. Additionally, this places the proposed method in the most significant category of $p$-values compared to other methods in the table.

\subsubsection{Comparison of Results on the HER2 Dataset}
We have evaluated the performance of the proposed method in comparison to other existing techniques, by using three separate criteria as outlined in the HER2 scoring contest \cite{qaiser2018her}, namely a) agreement points, b) weighted confidence (W.C) and c) combined points (C.Pts). Additionally, we incorporated Cohen's Kappa scores (C.K.), the percentage of average tissue area consulted while scoring a WSI, and the ROI selection process employed by each method. For the first assessment criterion of agreement points, upto 15 points were awarded based on the variation between the ground truth HER2 labels and the predicted HER2 scores for each case. The second assessment criterion, weighted confidence, was used to evaluate the reliability of the predictions by weighing each predicted score with a corresponding confidence value. The third criterion, combined points, was calculated as the product of agreement points and weighted confidence for each case. More details regarding assessment criteria are explained in \cite{qaiser2018her}. 

In order to calculate the agreement points for the proposed model, $N = 10$ tiles were extracted for each WSI and the most dominant class was selected as the slide-level score. Mean aggregation was used to aggregate the results of the four-folds to arrive at a final HER2 slide-level score for each WSI in the testing dataset. The confidence value was obtained by calculating the average probability values of the 10 tiles extracted for each WSI for each fold.

\begin{table*}[h]
\caption{HER2 Comparative Results: The table below presents the results of the proposed method, methods that competed in the HER2 Scoring Contest \cite{qaiser2018her} and post-challenge methods. The columns represent (in order) the agreement points, weighted confidence points (W.C), combined points (C.Pts), Cohen's Kappa scores (C.K.), the percentage of average tissue area (T.A.) consulted while scoring a WSI and the ROI selection process employed by each method.}
    \centering
    {\scriptsize
    \begin{tabular}{|l|c|c|c|c|c|c|}
    \hline
        \rule{0pt}{10pt} Methods & Points $\uparrow$ & W.C. $\uparrow$ & C.Pts $\uparrow$ & C.K. $\uparrow$ & \%T.A. $\downarrow$ & ROI Selection \\ \hline
        \textbf{The proposed method} & \textbf{402.5} & \textbf{23.59} & \textbf{347.2} &  \textbf{0.81} & \textbf{9.69\%} & Automated \\ \hline
        Learning Where to See \cite{qaiser2019learning} & \textbf{405} & \textbf{24.1} & \textbf{359.1} & \textbf{0.81} & 45.77\% & Sliding Window \\ \hline
        VISILAB-I (GoogleNet \cite{szegedy2015going}) & 382.5 & 23.55 & \textbf{348} & 0.76 & 100\% & Grid Technique \\ \hline
        VISILAB-II (contour analysis) & 377.5 & 21.88 & 322 & 0.66 & 45.77\% & Automated \\ \hline
        MTB NLP (AlexNet\cite{krizhevsky2012imagenet}) & 390 & 22.94 & 335.7 & 0.76 & 100\% & Manual \\ \hline
        Team Indus (LeNet\cite{lecun2015lenet}) & \textbf{402.5} & 18.45 & 321.4  & 0.61 & 45.77\%  & Sliding Window \\ \hline
        UC-CCSE \cite{mukundan2018image} & 390  & 21.07 & 316 & 0.66 & $\geq$ 11.97\%  & Semi-automated \\ \hline
        MUCS-III (AlexNet) \cite{pitkaaho2016classifying} & 390  & 20.43 & 300.8 & 0.56 & 100\% & Sliding Window \\ \hline
        HUANGCH (AdaBoost) & 377.5 & 22.62 & 345.7 & 0.71 & . & Automated \\ \hline
        MUCS-II (GoogleNet \cite{szegedy2015going}) & 385  & 19.51 & 290.1 & 0.56 & 100\% & Sliding Window \\ \hline
        FSUJena \cite{rodner2017deep} & 370 & 23 & 345 & 0.76 & . & Manual \\ \hline
        \multicolumn{7}{|c|}{Post-Challenge Methods}\\ \hline
        ZoomMil \cite{thandiackal2022differentiable} & 267.5 & 17.74 & 207.64 & 0.40 & \textbf{9.69\%} & Automated \\ \hline
        Attention based MIL \cite{ilse2018attention} & 355 & 11.4 & 143.50 & 0.48 & 100\% & Sliding Window \\ \hline
    \end{tabular}
    }
    \label{table:her2}
\end{table*}

Table \ref{table:her2} presents the agreement points, weighted confidence (W.C), combined points (C.Pts) and Cohen's Kappa scores (C.K.) across each method, along with the approximate percentage of average tissue area analyzed by each method and specifics of the ROI selection process. The proposed model, as shown in Table \ref{table:her2} achieved a total number of 402.5 points and a C.K score of 0.81 showing substantial agreement between the predicted and true labels. Our method outperformed majority of the other proposed approaches and achieved comparable performance to \cite{qaiser2019learning}, which extracted a total of approximately 58,000 image tiles, while our model achieved similar results while only extracting 510 tiles, with the help of attention mechanisms. 

Table \ref{table:her2} also shows the approximate percentage of average tissue area analyzed by each method. Methods that utilized $100\%$ of the tissue area employed tissue masks while methods that used $\approx46\%$ of the tissue area used a sliding window approach across the entire DAB region. In contrast, the proposed method, ZoomMil \cite{thandiackal2022differentiable} and UC-CCSE \cite{mukundan2018image} extracted image tiles only from informative regions. Our method analyzed approximately only $9.7\%$ of the tissue area using the automated soft attention-based ROI selection approach, achieving 402.5 points and a C.K. score of 0.81, significantly outperforming other methods that used a smaller tissue area.

\begin{table*}[]
\caption{HER2 Inference Time Results: The table presents the average time required for the inference phase for a whole slide image. Columns 'HA' and 'SA' indicate whether a model incorporates Hard Attention or Soft Attention, respectively, with a tick (\ding{51}) denoting usage and a cross (\ding{55}) indicating non-usage. Additionally, we present the number of parameters in each model.}
    \centering
    {\scriptsize
    \begin{tabular}{|c|c|c|c|c|}
    \hline
        \rule{0pt}{5pt} Model & SA & HA& Mean inference time per slide & Parameters \\ \hline
        \textbf{The proposed method} & \ding{51} & \ding{51} & \textbf{2.79} seconds & 3.09 M\\ \hline
        Learning Where to See \cite{qaiser2019learning} & \ding{55} & \ding{51} & 24.57 seconds & 3.08 M \\ \hline
        ResNet-34 \cite{he2016deep} & \ding{55} & \ding{55} & 12.62 seconds & 21.8 M \\ \hline
    \end{tabular}
    }
    \label{table:time}
\end{table*}

\subsubsection{Remarks on model efficiency}
To highlight the benefits of the dual attention method, we compare the inference time of the proposed model with other approaches on the HER2 Challenge Test Dataset. Table \ref{table:time} presents the average time (in seconds) required to process a WSI (for assigning a slide-level label). For each slide, the proposed model with the help of the soft attention module sampled $N = 10$ informative tiles which were subsequently processed by the hard attention module to arrive at a HER2 Score.  

We compare the proposed method which incorporates both hard and soft attention with the "Learning Where to See" model \cite{qaiser2019learning} which employs only hard attention and a ResNet-34 based model \cite{he2016deep}, a widely employed deep neural network (NN) architecture, effective for image processing tasks and commonly applied in CPath workflows. For this comparison, we extracted tiles of size 2048 $\times$ 2048 pixels (without overlap) using the sliding-window method after the application of DAB masks and processed them with each of the aforementioned methods.  

As shown in Table \ref{table:time} the proposed method which employs both soft and hard attention, significantly reduces the time needed to evaluate a WSI. The proposed model improves inference time by approximately $88\%$ compared to "Learning Where to See" \cite{qaiser2019learning} and by approximately $77\%$ compared to ResNet-34. This efficiency is achieved by selectively extracting informative tiles from regions of interest, as identified by the soft attention module, resulting in a considerable reduction in the time required to process and classify a WSI. 

We also observe that the proposed model, which incorporates both soft and hard attention mechanisms, introduces a marginal increase in the number of parameters compared to the model that employs hard attention alone \cite{qaiser2019learning}. Despite this enhancement in functionality, the parameter count remains significantly lower than that of the ResNet-34 model \cite{he2016deep} . It is also important to note that the proposed model operates of the WSI-level while the other two methods operate on a tile-level. This highlights the effectiveness of our approach in integrating both attention mechanisms with minimal computational overhead. 

\begin{figure*}[!t]
\centerline{\includegraphics[width=0.9\textwidth]{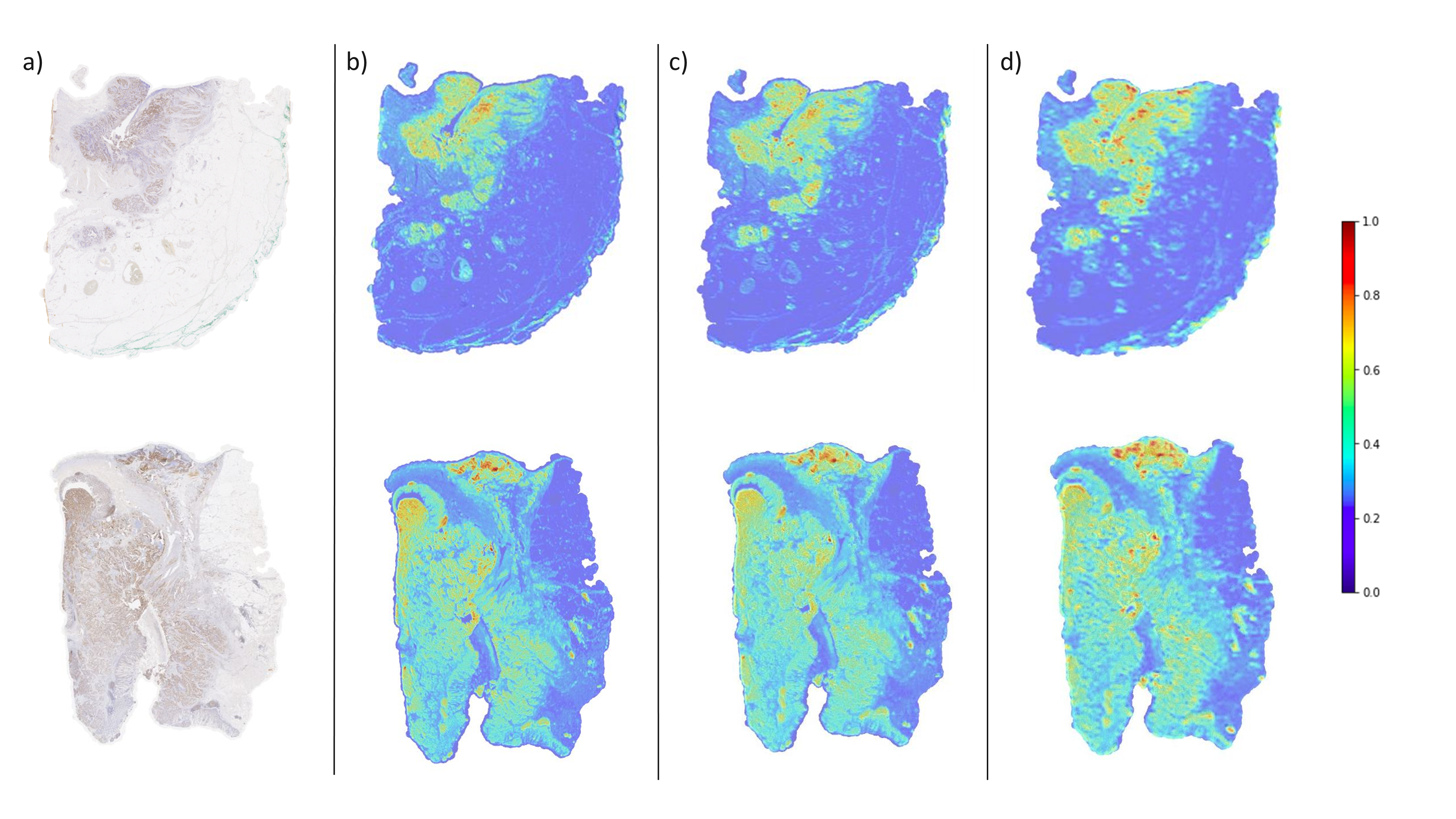}}
\caption{a) Original images; b) Attention maps generated at magnification 1.25$\times$; c) Attention maps generated at magnification 0.625$\times$; d) Attention maps generated at magnification 0.3125$\times$. The first row shows a sample from PMS2 while the second row shows a sample from MLH1.} 
\label{fig_att_mags}
\end{figure*}  

\subsection{Influence of number of image tiles on performance}
We have conducted an additional study to evaluate the performance of the dual attention model with respect to the number of tiles. We selected $N = [10,15, 20]$ tiles per WSI for this study. From each image tile, six multi-resolution glimpses or patches were extracted to process the given tile. The experiments were run on the PMS2 test set and the results were averaged over 10 runs to account for the noise introduced during the attention sampling procedure and to reduce the impact of the inherent stochastic nature of the RL hard attention classifier. The mean, mode and variance of the independent runs on the test set with varying numbers of image tiles are shown below to exhibit the accuracy and stability of the proposed model. We have observed that the performance plateaus after reaching a certain number of tiles. Additionally, increasing the number of tiles leads to an overhead in $6 \times$ processing time, as each additional image tile results in an additional six glimpses per tile. 
We observe from Table. \ref{table:tiles} that for Class 0 the performance plateaus at 15 tiles. Additionally, the number of image tiles must not be so high that the model extracts overlapping tiles from low attention areas. In this regard, we opted for a middle ground by choosing $N=15$ for our experiments for the COMET-MMR dataset and  $N=10$ for HER2.


\begin{table}[]
\caption{Class-wise Accuracy of the Dual-Attention Model on PMS2 Testing Dataset with Varying Numbers of Tiles}
\centering

\begin{tabular}{|c|c|c|}
 \hline
 \multicolumn{3}{|c|}{PMS2 Testing Data set} \\
 \hline
 Label & 0  & 1  \\
 \hline
 No.of Tiles & Class-wise Accuracy & Class-wise Accuracy \\
 \hline
 10 & $0.53 \pm 0.22 $ & $0.68 \pm 0.11$ \\ \hline
 15 & $ 0.67 \pm 0.001 $ & $0.70 \pm 0.04$ \\ \hline
 20 & $ 0.67 \pm 0.14 $ & $0.77 \pm 0.05$ \\ 
 \hline
\end{tabular}

\label{table:tiles}
\end{table}

\begin{figure*}[!t]
\centering
\includegraphics[page=1, width=0.8\textwidth]{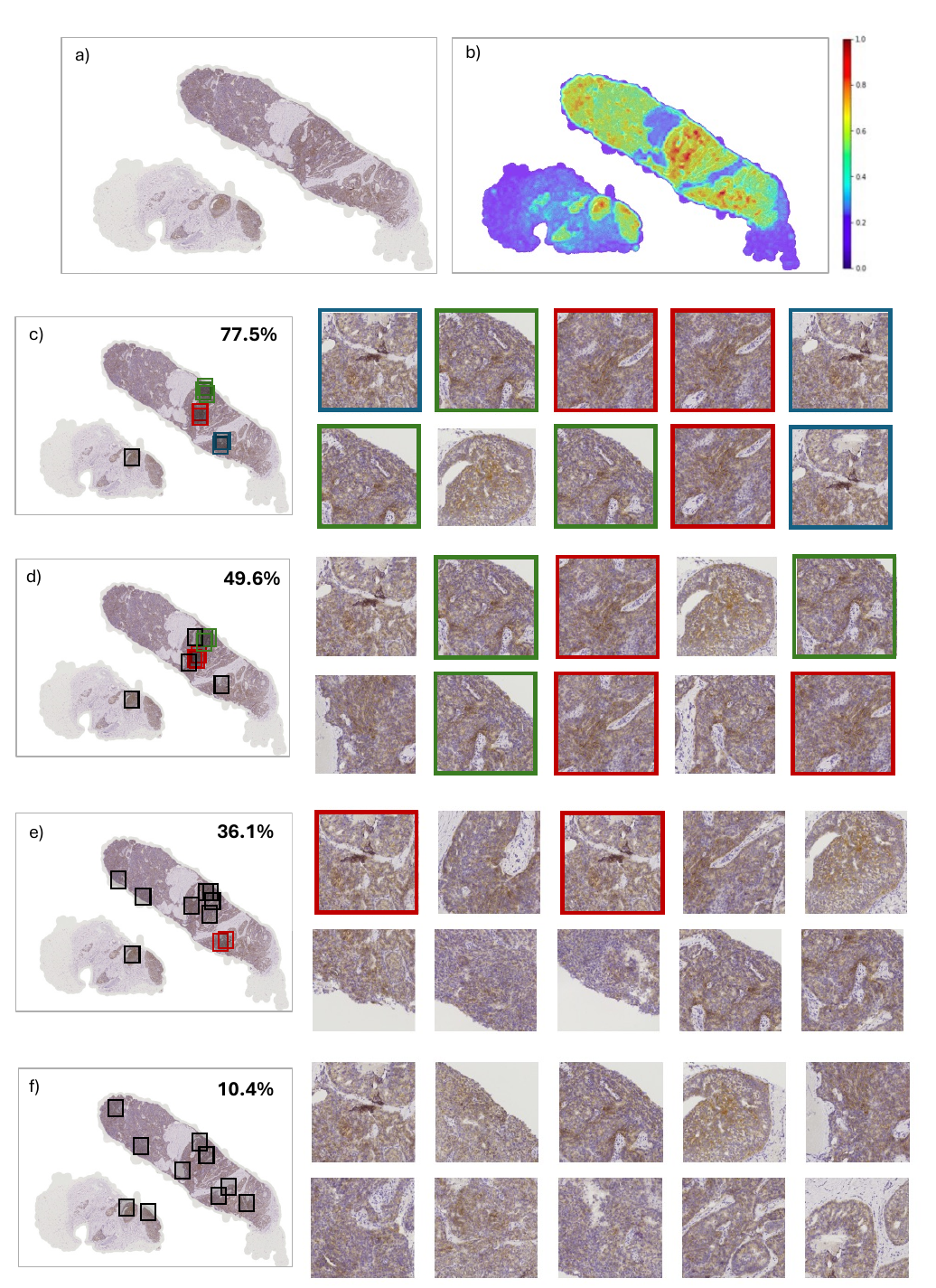}
\caption{(left to right) a) Original image ; b) Attention map ; c) The locations of the tiles and tiles extracted only using top-$k$ sampling ; d) The locations of the tiles and tiles extracted when performing the distance-constrained point selection step (DCPS) ; e) The locations of the tiles and tiles extracted when noise is added to the sampling process; f) The locations of the tiles and tiles extracted using the proposed method (noise + DCPS). The clusters of overlapping tiles are highlighted using different colors. }
\label{fig_noise}
\end{figure*}

\begin{figure*}[!t]
\centering
\includegraphics[page=1, width=\textwidth]{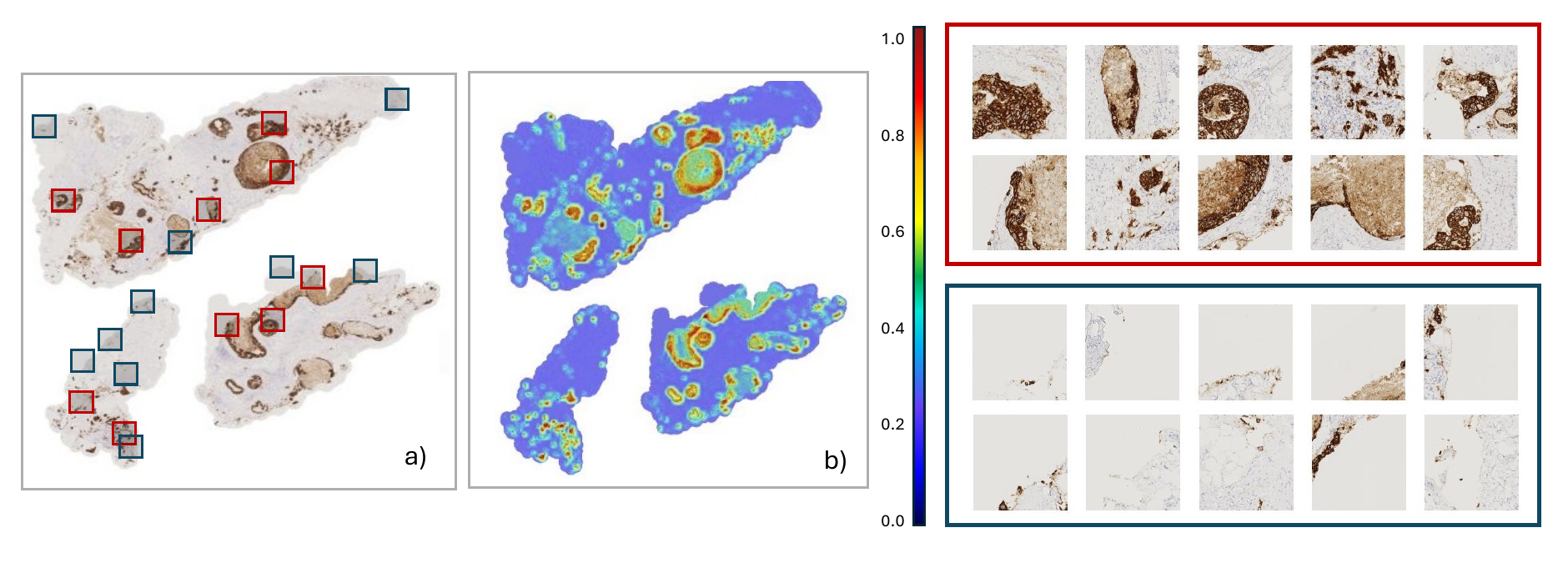}
\caption{ a) The original image with marked locations of tiles extracted based on attention levels; b) the corresponding attention map.  Tiles from high attention regions are indicated with red outlines, while those from low attention regions are marked in blue. High magnification views of these tiles are displayed within corresponding red and blue boxes. }
\label{high_low_attention}
\end{figure*}

\subsection{Qualitative results}
\subsubsection{Influence of magnification on performance}
We have conducted experiments to evaluate the sensitivity of the soft attention block with respect to the magnification level used to train the model. In the proposed workflow the WSIs were down-sampled by a factor of 32 to level 5 at magnification 1.25$ \times$ to train the soft attention module. We have conducted additional experiments by training the soft attention module using images extracted at magnifications 0.625 $\times$ (level 6) and 0.3125 $\times$ ( level 7) for the two Comet-MMR markers, MLH1 and PMS2. In Fig. \ref{fig_att_mags} we see that the model is able to effectively capture and generalize to the relevant features and structures present in the images across different magnifications.

\subsubsection{Influence of the attention sampling procedure}
The proposed two-step sampling scheme introduces diversity in the choice of informative tiles provided to the classifier during the training process. In Fig. \ref{fig_noise} we show that for a given slide employing only top-$k$ for the sampling of image tiles results in multiple clusters of overlapping tiles in the high attention regions as highlighted in the attention map. The percentage of overlap in the test set when using only top-$k$ sampling is 77\% on average. The addition of the DCPS strategy and targeted noise, when used independently reduces the number of overlapping tiles to $49\%$ and $36\%$ on average, as can be seen in the figure. However, using both methods together significantly improves the chances of selecting a set of diverse and informative image tiles and reduces the amount of redundant information passed to the RL classification model. Fig. \ref{fig_noise} demonstrates that the utilization of both methods results in a reduction of overlap to $10\%$ equating to one in every 10 sampled tiles on average. This sampling method therefore, significantly enhances the representative ability of the sampled tiles, and consequently improves the reliability of the classification label.

We have also sampled image tiles from low attention areas by adjusting the threshold accordingly to show that the proposed method accurately distinguishes high attention areas even in WSIs with sparse staining . It can be seen from Fig. \ref{high_low_attention} that the tiles sampled from high attention area have higher intensity of IHC staining whereas the tiles sampled from the lower attention areas have not only lower intensity of IHC staining but image tiles extracted from the non-tumor and background regions as well. 

\section{Conclusions and Discussion}
In this study, we have presented a novel dual hierarchical attention model that combines soft attention at the slide-level and hard attention at the patch-level for the automated scoring of IHC stained WSIs. We derive inspiration from a pathologist's visual analysis of a tissue slide, by first employing a soft attention mechanism at low resolution to identify areas in need of further investigation and then extracting a set of spatially distinct and diverse image tiles from these regions. Additionally, instead of processing these extracted tiles entirely, we further employ reinforcement learning to extract a sequence of multi-resolution patches to score each tile. These scores are then aggregated to obtain a slide-level classification decision. We trained the two attention modules in an end-to-end manner for two WSI-level classification tasks, predicting the binary Loss/Intact status for MLH1 and PMS2 biomarkers in colorectal cancer and HER2 scoring in breast cancer. This is achieved by combining their respective loss functions into a joint dynamic loss function, which is optimized during training.


Most WSI pre-processing pipelines employ the sliding window paradigm, where the standard approach is to independently extract and save the image patches. Not only is this a time-consuming task but it requires vast reserves of memory. We take inspiration from the principle of “working smarter, not harder”, and achieve results better or comparable to the state-of-the-art methods while processing a fraction of the WSI at the highest resolution. While the average WSI can be of the order of $10^{10}$ pixels \cite{zhou2021histopathology}, we perform slide-level classification while only attending to 6 glimpses of size 128 $\times$ 128 each, at the highest resolution (40$\times$ and 20$\times$) from each of the approximately 20 image tiles extracted for single slide, which amounts to less than 5\% of the entire slide area and less than 10\% of the tissue area, while achieving results better or comparable to state-of-the-art methods.

Additionally, the two-step tile sampling scheme, which includes the injection of target noise and the distance-constrained point selection step, has enabled us to extract a spatially diverse and informative set of image tiles. This approach has reduced the amount of overlap by approximately $60\%$, allowing us to capture a better overall representation of the whole slide image for downstream purposes.


Moreover, processing only a small set of representative tiles instead of analyzing all tiles within a WSI can significantly reduce the computational time required to assign slide-level labels to the WSIs. The proposed method is able to assign a slide-level classification label while only extracting a small set of representative tiles resulting in an average processing time of approximately 3 seconds for each slide, thereby improving inference times by approximately $77 \%$ and  $88 \%$ compared to CNN and hard attention attention based methods. 

The proposed method also attempts to address the memory issues inherent in all patch-based models, by extracting a batch of image tiles in real time, performing the required analysis, updating the gradients accordingly, and then discarding the image tiles without saving them, potentially saving memory. This can allow for the deployment of this method in labs with limited memory resources. Additionally, in standard WSI pipelines which extract hundreds of image tiles per slide, it is difficult to ascertain which tiles have contributed to the final score. In comparison, our selective attention mechanism which identifies truly useful image tiles, can allow a pathologist to conveniently assess and corroborate the predicted results and can be incorporated as an assistive diagnostic tool.

A key limitation in this study is determining the optimal number of tiles that should be extracted from a given WSI. Although the goal is to obtain a representative set of image tiles to accurately classify the WSI, the exact number required to capture all the essential information within a given WSI remains uncertain. Future work for this study includes exploring approaches for determining the optimal tile selection criteria and adapting the number of tiles extracted to the specific characteristics and tissue area of each individual WSI.





\bibliographystyle{elsarticle-num} 
\bibliography{main}






\end{document}